\documentstyle[aps,twocolumn,10pt]{revtex}
\newlength{\defaultparindent}
\newenvironment{Default Paragraph Font}{}{}
\begin{document}

\title{Two-vibron bound states lifetime in a one-dimensional molecular lattice coupled to acoustic phonons}

\author{V. Pouthier}
\address{Laboratoire de Physique Mol\'{e}culaire, UMR CNRS 6624. Facult\'{e} des
Sciences - La Bouloie, \\ Universit\'{e} de Franche-Comt\'{e}, 25030 Besan\c {c}on cedex, 
France.}
\date{\today}
\maketitle

\begin{abstract}
The lifetime of two-vibron bound states in the overtone region of a one-dimensional anharmonic molecular lattice is investigated. The anharmonicity, introduced within an attractive Hubbard Hamiltonian for bosons, is responsible for the formation of bound states which belong to a finite linewidth band located below the continuum of two-vibron free states. The decay of these bound states into either bound or free states, is described by considering the coupling between the vibrons and a thermal bath formed by a set of low frequency acoustic phonons. The relaxation rate is expressed in terms of the spectral distribution of the vibron/phonon coupling and of the two-vibron Green operator which is calculated exactly by using the number states method. The behavior of the two-vibron bound states relaxation rate is analyzed with a special emphasis on the influence of the anharmonicity. It is shown that the rate exhibits two distinct regimes depending on the thermal bath dimension. When the bath dimension is equal to unity, the rate increases with the anharmonicity and the decay of the two-vibron bound states into the other bound states appears as the main contribution of the rate. By contrast, when the bath dimension is equal to 2 and 3, the rate decreases as the anharmonicity increases indicating that the two-vibron bound states decay into the two-vibron free states continuum.    
\end{abstract}

\section{Introduction}

The nonlinear nature of vibrational excitations in low-dimensional molecular lattices 
plays a key role for energy transfer as well as energy storage in both physical, 
chemical and biological systems. 

This feature was first pointed out by Davydov and co-workers \cite{kn:davydov} to 
explain how the energy released by the hydrolysis of ATP, partially stored in the 
high frequency C=O vibration (amide-I) of a peptide group of a protein, can be 
transported from one end of the protein to the other. The dipole-dipole coupling between the different peptide groups leads to the delocalization of the internal vibrations 
and to the formation of vibrons. The interaction between the vibrons and the 
phonons of the protein induces a nonlinear dynamics which counterbalances the dispersion 
created by the dipole-dipole coupling. It yields the creation of the so called Davydov's 
soliton which appears as the solution of the Nonlinear Schrodinger (NLS) equation  within 
the continuum approximation. Nevertheless, it has been shown that several properties, such as the space dimension \cite{kn:kuznetsov} and the discreteness of the lattice \cite{kn:brizhik}, 
tend to destroy the integrability of the NLS equation. 

However, a remarkable feature given by the discrete version of this equation is the occurrence of intrinsic localized modes or discrete breathers (for a recent review, see for instance Ref. \cite{kn:aubry}). Contrary to solitons, intrinsic localized modes do not require integrability for their existence and stability. They are not restricted to one-dimensional lattices and it has been suggested that they should correspond to quite general and robust solutions \cite{kn:takeno}. In molecular lattices, discrete breathers can be induced by the 
intramolecular anharmonicity of the molecules. Indeed, within the semi-classical approximation \cite{kn:takeno1}, the vibron dynamics is described by the discrete NLS equation. 
As a result, the presence of an intrinsic localized mode corresponds to a local accumulation 
of the internal energy, which might be pinned in the lattice or may travel through it. Breather excitations are thus expected to be of fundamental importance for both vibrational energy storage and transport in molecular lattices. 

In spite of the great interest that breathers have attracted, no clear evidence has yet been found for their existence in real molecular lattices. By contrast, two-vibron bound states (TVBS), which can be viewed as quantum breathers, have been observed in several systems. Indeed, in quantum lattices, the intramolecular anharmonicity, which is responsible for the interaction between the vibrons, favors the formation of bound states \cite{kn:kimball}. When two vibrons are excited, a bound state corresponds to the trapping of two quanta over only a few neighboring molecules with a resulting energy which is lesser than the energy of two quanta lying far apart. The lateral interaction yields a motion of such a state from one molecule to another, thus leading to the formation of a delocalized wave packet with a well-defined momentum. As mentioned by Scott\cite{kn:scott}, these bound states can thus be viewed as the quantum counterpart of breathers or soliton excitations. From an experimental point of view, the formation of TVBS in low-dimensional molecular lattices was observed in molecular adsorbates such as H/Si(111) \cite{kn:guyot1}, H/C(111) \cite{kn:shen}, CO/NaCl(100) \cite{kn:dai} and CO/Ru(100) \cite{kn:jakob1} using optical probes. Bound states in the system H/Ni(111) were investigated using high resolution electron energy loss spectroscopy \cite{kn:okuyama}. In addition, the powerfulness of 
two-vibron infrared spectroscopy in probing the structure of admonolayers has been 
recently demonstrated both experimentally \cite{kn:jakob2} and theoretically \cite{kn:pouthier1}.  

The stability of both classical and quantum breathers under the influence of a fluctuating thermal bath is of fundamental importance for energy storage and energy transport at finite 
temperature. For instance, at biological temperature, the lifetime of the Davydov's soliton 
is still an open question. Indeed, it has been shown that the amide-I excitation, in vivo, corresponds to a localized state \cite{kn:cruzeiro1,kn:cruzeiro2}. Instead of traveling in a coherent manner, it follows a stochastic, diffusional path along the lattice. In other words, the single-vibron Davydov soliton does not last long enough to be useful at biological temperatures and multi-vibron solitons have been proposed as good candidates \cite{kn:tekic,kn:pang}.
In the same way, the influence of white noise and nonlinear damping on the stability of the breathers of the discrete NLS equation has been studied \cite{kn:christiansen}. It has been shown that the lifetime of the breathers increases as the nonlinearity increases 
and is inversely proportional to the noise variance. By contrast, for quantum lattices, it was pointed out by Jakob et al. \cite{kn:jakob3} that the dephasing constant of TVBS is larger than the single-vibron dephasing constant by a factor four. This feature, which occurs when both kinds of states interact in the same way with the thermal bath, leads to a decrease of the TVBS lifetime when compared with the single-vibron lifetime. Similar results were previously obtained by Bogani et al. \cite{kn:bogani} who have characterized the relaxation of bound states in the combination regions of two internal modes in a molecular crystal. The authors have shown that the lifetime of a bound state decreases as the anharmonicity increases, excepted at very low temperature. They assumed that the anharmonicity yields the occurrence of a bound state with zero linewidth and that the relaxation, mediated by the phonons of the crystal, leads to the decay of the bound state into the two-vibron continuum.

In this paper, a theoretical study is performed to investigate the TVBS lifetime in the overtone region of a one-dimensional molecular lattice. The bound states, which are described by using a local approach, belong to a finite linewidth energy band located below the continuum of two-vibron free states. The relaxation processes are considered by introducing a coupling between the vibrons and a thermal bath formed by a set of low frequency phonons. Such a coupling, which is assumed to be responsible for dephasing 
mechanism only, does not allow for energy relaxation. In other words, the interaction with the thermal bath does not modify the vibron number but induces transitions between the different two-vibron eigenstates. The formalism can be applied to describe different physical 
situations such as the dynamics of low-dimensional adsorbed nanostructures (molecular nanowire adsorbed on stepped surfaces) and the collective vibration of the peptide groups in a protein (vibrons in $\alpha$-helix polypeptide proteins). In the first situation, the thermal bath is formed by the phonons and the librons of the adsorbed nanowire and by the phonons of the substrate \cite{kn:pouthier2,kn:pouthier3,kn:pouthier4}. In the second situation, it is characterized by the low frequency vibrational modes of the protein as well as by the 
biological surrounding. 

The paper is organized as follows. In Sec. II, the Hamiltonian of the vibrons and of the thermal bath, as well as the coupling Hamiltonian, are described. In Sec. III, we first summerize the number states method used to characterize the TVBS and the equivalence between the two-vibron dynamics and the motion of a single fictitious particle in a one-dimensional lattice is established. Then, the TVBS relaxation rate due to the coupling with acoustic phonons is expressed in terms of the two-vibron Green operator which is finally determined.
In Sec. IV, a detailed analysis of the behavior of the rate is performed depending on the values taken by the relevant parameters of the problem (the anharmonicity, the temperature, the thermal bath dimension). These results are finally discussed in Sec. V.
 
\section{Model and Hamiltonian}

Let us consider a set of $N$ molecules periodically distributed on the sites 
of a one-dimensional lattice. To characterize the vibron dynamics of this 
molecular lattice, we assume that each molecule $n$ behaves as an internal high frequency oscillator described by the standard creation and annihilation vibron operators $b^{+}_{n}$ and $b_{n}$. The Hamiltonian which describes the collective dynamics of the internal vibrations is expressed as (using the convention $\hbar=1$)
\begin{equation}
H_{v}= \sum_{n} h_{n}+ \sum_{n,n'} \Phi(n,n') b^{+}_{n}b_{n'}
\end{equation}
where $h_{n}$ defines the internal vibration Hamiltonian of the $n$th molecule
and $\Phi(n,n')$ represents the lateral dynamical coupling between the internal 
coordinates at the quadratic order. To minimize the number of parameters, we consider interactions between nearest neighbor molecules and define the hopping constant 
$\Phi$. Note that long range lateral interactions may affect the two-vibron dynamics by inducing changes in the vibron dispersion curve. However, these effects are expected to be rather weak for a one-dimensional lattice and are disregarded in the present 
work.
The internal vibration Hamiltonian $h_{n}$, including the intramolecular anharmonicity of each molecule, is described according to the model of Kimball et al. \cite{kn:kimball}. In this model, the intramolecular potential is expanded up to the fourth order with respect to the internal coordinate and a unitary transformation is performed to keep the vibron-conserving terms, only.  The resulting Hamiltonian $H_{v}$ is essentially a Bose version of the Hubbard model with attractive interactions written as,   
\begin{equation}
H_{v}=\sum_{n} \omega_{0}b^{+}_{n}b_{n}-Ab^{+}_{n}b^{+}_{n}b_{n}b_{n}
+ \Phi (b^{+}_{n}b_{n+1}+b^{+}_{n+1}b_{n})
\label{eq:Hv}
\end{equation}
where $\omega_{0}$ stands for the internal frequency of the $n$th molecule and $A$ denotes the positive anharmonic parameter.

These molecules interact with their surrounding which acts as a thermal bath and which is responsible for the relaxation mechanism. To mimic the influence of the surrounding of the molecules, we assume that the thermal bath is formed by a set of low frequency harmonic oscillators which the Hamiltonian is expressed as 
\begin{equation}
H_{b}=\sum_{q} \Omega_{q}a^{+}_{q}a_{q}
\label{eq:Hb}
\end{equation}
where $a^{+}_{q}$ and $a_{q}$ stand for the boson operators of the $q$ th mode with frequency $\Omega_{q}$. By neglecting the population relaxation, the thermal bath leads to stochastic fluctuations of the dynamical parameters (internal frequency, anharmonic parameter, force constant) which characterize the vibron dynamics. In this work, 
we consider that the main effect of the bath is to induce a random modulation of the internal frequency of each molecule. The coupling Hamiltonian between the vibrons and the thermal bath is thus written as
\begin{equation}
\Delta H_{vb}=\sum_{n} \Delta\omega_{n}b^{+}_{n}b_{n}
\label{eq:DeltaH}
\end{equation}
where $\Delta\omega_{n}$ is an operator with respect to the bath variables. In a general way, it depends on the bath coordinates in a very complicated manner. However, to simplify our discussion we consider the linear approximation in which $\Delta\omega_{n}$ is expressed as
\begin{equation}
\Delta \omega_{n}=\sum_{q} \chi_{q,n} a^{+}_{q}+\chi^{*}_{q,n} a_{q}
\label{eq:Deltaom}
\end{equation}
The previous model has been extensively used to study relaxation mechanisms in condensed matter (see for instance Refs. \cite{kn:dibartolo,kn:skinner,kn:berne,kn:golosov}) and it has been shown that the effect of the thermal bath is completely determined by the coupling distribution $J(\omega)$ defined as
\begin{equation}
J(\omega)=\frac{1}{N} \sum_{n,q} \mid \chi_{q,n} \mid^{2} \delta(\omega-\Omega_{q})
\label{eq:J1}
\end{equation}

The theory presented below to calculate the TVBS relaxation rate depends on the coupling distribution and can thus be applied to different kinds of thermal baths. In the present work, we shall restrict our analysis to an interaction with a set of acoustic phonons. In that case, the coupling distribution is determined by assuming a Debye model for the phonon density of states coupled with a deformation potential approximation \cite{kn:dibartolo}. Therefore, $J(\omega)$ scales as $\omega^{\alpha}$, where $\alpha$ stands for the dimension of the phonon lattices, and is defined as, 
\begin{equation}
J(\omega)= \left\{ \begin{array}{ll}
J_{0}(\omega / \Omega_{c})^{\alpha} & \mbox{if $\omega < \Omega_{c}$} \\
0                                   & \mbox{if $\omega > \Omega_{c}$}
\end{array}
\right.
\label{eq:J2}
\end{equation}
where $\Omega_{c}$ denotes the cutoff frequency of the phonon bath. Note that, in the present paper, the bath dimension is considered as a parameter which can take the values 1, 2 and 3 in order to represent different physical situations. For instance, for an adsorbed one-dimensional molecular lattice, the vibrons can be coupled preferentially with the phonons of the molecular lattice. The resulting bath dimension is thus equal to one. In the same way, for a molecular adsorbate decorating a surface step, the vibrons can be coupled with the step phonons which exhibit a one-dimensional nature. Moreover, we can imagine other situations such as a strong coupling with either surface phonons or bulk phonons for which the bath dimension is equal to two or three, respectively.

\section{Theoretical background}

\subsection{Two-vibron states}

In this section, the number states method \cite{kn:scott} is summarized in order
to characterize the two-vibron states of the molecular lattice. 
Within this method, the eigenstates of the Hamiltonian $H_{v}$ (Eq.(\ref{eq:Hv})) are determined by taking advantage of the fact that $H_{v}$ commutes with the operator $\sum_{n} b_{n}^{+}b_{n}$ which counts the total number of vibrational quanta. As a result, the Hilbert space 
$E$ of the vibrational states can be written as the tensor product $E= E_{0} \otimes E_{1} \otimes E_{2} \otimes ... \otimes E_{p} ...$, where $E_{p}$ denotes the subspace 
connected to the presence of $p$ vibrational quanta. Single-vibron states, connected to the subspace $E_{1}$ of dimension $N$, correspond to plane waves with wave vectors $q$ lying inside the first Brillouin zone of the chain and with eigenfrequencies $\omega(q)=\omega_{0}+2\Phi cos(q)$.

Two-vibron states are related to the subspace $E_{2}$, which the dimension 
$N(N+1)/2$ represents the number of ways for distributing two indistinguishable quanta onto $N$ sites. 
A useful basis set to generate the $E_{2}$ subspace can be constructed by applying 
two creation operators on the vacuum state $\mid 0 \rangle$ as   
\begin{equation}
\mid n_{1},n_{2} ) = 
[1-\delta_{n_{1}n_{2}}(1-\frac{1}{\sqrt{2}})]
b_{n_{1}}^{+}b_{n_{2}}^{+} \mid 0 \rangle
\label{eq:basis}
\end{equation}
where the factor in the right hand side of Eq.(\ref{eq:basis}) is a normalization coefficient and where the restriction $n_{2} \geq n_{1}$ is applied to avoid counting twice the vectors ${\mid n_{1},n_{2})}$ due to the indistinguishable nature of the quanta.
As a result, to solve the Schrodinger equation $H_{v} \mid \Psi \rangle = \omega \mid \Psi \rangle$, the most general two-vibron eigenstate can be expanded as a linear combination 
of the previous basis vectors, as
\begin{equation}
\mid \Psi \rangle = \sum_{n_{1},n_{2}} \Psi(n_{1},n_{2}) \mid n_{1},n_{2} )
\end{equation}
However, by taking advantage of the  
lattice periodicity, an intermediate basis set can be determined. Indeed, since the 
two-vibron wave function $\Psi(n_{1},n_{2})$ is invariant with respect to a translation 
along the lattice, it depends on $n_{1}$ and $m=n_{2}-n_{1}$, and it can be 
written as a Bloch wave, as
\begin{equation}
\Psi(n_{1},n_{2}) = \frac{1}{\sqrt{N}} \sum_{k} \Psi_{k}(m) e^{ik(n_{1}+n_{2})/2}
\label{eq:bloch}
\end{equation}
The total momentum $k$ of the two-vibron states is associated to the motion of the center of mass of the two quanta. The values of $k$ are determined by imposing periodic boundary conditions to the wave function over a set of $N$ unit cells.
The resulting wave function $\Psi_{k}(m)$ refers to the degree of freedom $m$ which characterizes the reduced distance between the two vibronic excitations. It corresponds 
to the projection of the eigenstate $\mid \Psi \rangle$ on the basis set $\mid k,m \rangle$
defined as,
\begin{equation}
\mid k,m \rangle = \frac{1}{\sqrt{N}} \sum_{n_{1}} e^{ik(n_{1}+m/2)}\mid n_{1},n_{1}+m )
\label{eq:bloch1}
\end{equation}
Note that the previous periodic boundary conditions lead to limit the values taken by the 
index $m$ which cannot exceed the half size of the periodic box and thus varies from 0 to 
$(N-1)/2$. Moreover, specific boundary conditions are introduced depending on the parity of the number $N$ \cite{kn:scott}. However, by considering the limit $N \rightarrow \infty$, these features will be disregarded in this work.

Since the total momentum $k$ is a good quantum number, the Hamiltonian $H_{v}$ appears as
block diagonal in the intermediate basis $\mid k,m \rangle$ (Eq.(\ref{eq:bloch1})) and the Schrodinger equation can be solved for each $k$ value. For a given $k$ value, the representation of the restricted Hamiltonian $H_{vk}$ depends on the nature of the basis vectors involved in. Indeed, there are two different kinds of intermediate basis vectors either describing two quanta located on molecules which are far apart ($m > 1$) or two quanta located onto the same molecule ($m=0$). In the first situation, the Schrodinger equation is expressed as 
\begin{equation}
(2\omega_{0}-\omega)\Psi_{k}(m) +\tilde{\Phi}_{k}(\Psi_{k}(m+1)+\Psi_{k}(m-1)) = 0
\label{eq:schrod1}
\end{equation}
where $m>1$ and $\tilde{\Phi}_{k}=2\Phi cos(k/2)$. When $m=1$, 
the Schrodinger equation becomes, 
\begin{equation}
(2\omega_{0}-\omega)\Psi_{k}(1)+\tilde{\Phi}_{k}\Psi_{k}(2)
+\sqrt{2}\tilde{\Phi}_{k}\Psi_{k}(0)=0
\label{eq:schrod2}
\end{equation}
whereas when m=0, it is written as 
\begin{equation}
(2\omega_{0}-2A-\omega)\Psi_{k}(0) + \sqrt{2}\tilde{\Phi}_{k} \Psi_{k}(1)=0
\label{eq:schrod3}
\end{equation}

Eqs.(\ref{eq:schrod1})-(\ref{eq:schrod3}) clearly indicate how the physics of the two-vibron states is related to the dynamics of a fictitious single particle moving quantum mechanically in a one-dimensional lattice \cite{kn:kimball}. This lattice, shown in Fig. 1a, appears as a discrete, finite size chain with a length which extends from $m=0$ to $(N-1)/2$. 
Within this equivalence, the wave function $\Psi_{k}(m)$ associated to the relative distance between the two vibrons, can be viewed as the wave function of the fictitious particle. Its dynamics is described by a tight-binding Hamiltonian $H_{vk}$ characterized by the self-energy $2\omega_{0}$ located on each site and a hopping matrix $\tilde{\Phi}_{k}$ which couples nearest neighbor sites. The anharmonicity is responsible for the presence of a local defect on the site $m=0$ which induces a redshift $-2A$ of the corresponding self-energy. This defect allows us to discriminate between two different eigenstates. The eigenstates of the core of the chain correspond to plane waves slightly perturbed by the presence of the defect and are connected to the free propagation of the fictitious particle. In terms of two-vibron states,  these eigenstates correspond to a delocalization of the separating distance between the two quanta and leads to the occurrence of two-vibron free states (TVFS) which belong to an energy band located around $2\omega_{0}$ (Fig. 2a). Note that if there is no anharmonicity ($A=0$), a TVFS with a total momentum $k=q_{1}+q_{2}$ corresponds to the presence of two independent vibrons with wave vector $q_{1}$ and $q_{2}$ and energy $\omega(q_{1})+\omega(q_{2})$.

By contrast, the presence of a defect is responsible for the occurrence of a localized state for the fictitious particle. This state is associated to an exponential decrease of the wave function from the side $m=0$ of the chain. In terms of two-vibron states, it corresponds to a TVBS which the wave function is defined as \cite{kn:kimball,kn:scott}
\begin{equation}
\Psi^{B}_{k}(m)=\sqrt{\frac{4A}{2\omega_{0}-\omega_{B}(k)}}
(1+\frac{1-\sqrt{2}}{\sqrt{2}}\delta_{m,0})e^{-m/\xi_{k}}
\label{eq:TVBSWF}
\end{equation}
where $\omega_{B}(k)$ stands for the TVBS frequency written as
\begin{equation}
\omega_{B}(k)=2\omega_{0}-2\sqrt{A^{2}+4\Phi^{2}cos^{2}(k/2)}
\label{eq:TVBSE}
\end{equation}
and where $\xi_{k}$ is the length of the bond between the two vibrons expressed as
\begin{equation}
\xi^{-1}_{k}=ln \sqrt{\frac{2\omega_{0}-\omega_{B}(k)+2A}{2\omega_{0}-\omega_{B}(k)-2A}}
\label{eq:xi}
\end{equation}
As shown in Fig. 2a, TVBS belong to a low frequency band located below the TVFS continuum and characterized by the dispersion curve $\omega_{B}(k)$. 

\subsection{Two-vibron bound state relaxation rate} 

Due to the coupling Hamiltonian $\Delta H_{vb}$ (Eq.(\ref{eq:DeltaH})), TVBS do not represent exact eigenstates of the whole system "vibron+phonon bath". More precisely, this coupling is responsible for the occurrence of transitions between two-vibron states mediated by the emission or the absorption of acoustic phonons.  

By assuming the coupling Hamiltonian $\Delta H_{vb}$ as a weak perturbation, the rate for the transition from a TVBS $\mid \Psi^{B}_{k} \rangle$ to another two-vibron state $\mid \Psi^{\sigma}_{k'} \rangle$\ (bound $\sigma=B$ or free $\sigma=F$ states) with frequency $\omega_{\sigma}(k')$ can be expressed by using the Fermi Golden Rule as 
\begin{eqnarray}
W(\Psi^{B}_{k}\rightarrow \Psi^{\sigma}_{k'})&=& 2 \pi
\sum_{\alpha,\beta} P_{\alpha} \mid \langle \Psi^{B}_{k},\alpha \mid \Delta H_{vb} \mid
\Psi^{\sigma}_{k'},\beta \rangle \mid^{2} \nonumber \\ 
& &\delta(\omega_{B}(k)+\Omega_{\alpha}-\omega_{\sigma}(k')-\Omega_{\beta})  
\label{eq:rate1}
\end{eqnarray}
Eq.(\ref{eq:rate1}) describes a transition in the course of which the thermal bath evolves from an initial state $\mid \alpha \rangle$ with frequency $\Omega_{\alpha}$ to a final state $\mid \beta \rangle$ with frequency $\Omega_{\beta}$. Since the thermal bath is assumed to be in thermal equilibrium at temperature $T$, a statistics is realized over the initial state with the probability occupation $P_{\alpha}$ and a sum  over all the possible final bath states is performed. By inserting the expression of the coupling Hamiltonian $\Delta H_{vb}$ (Eq.(\ref{eq:DeltaH})), the total rate for leaving the bound state $\mid \Psi^{B}_{k} \rangle$ obtained by summing over all possible transitions is expressed as
\begin{eqnarray}
& &W^{(2)}_{k,B} = 2Re \sum_{n,n'} \int_{0}^{\infty} dt \langle \Delta \omega_{n}(t)
\Delta \omega_{n'}(0) \rangle  \nonumber \\ 
& & \langle \Psi_{k}^{B} \mid b_{n}^{+}(t) b_{n}(t)b_{n'}^{+}(0) b_{n'}(0) \mid  \Psi_{k}^{B} \rangle
\label{eq:rate2}
\end{eqnarray}
where $\langle ... \rangle$ stands for an average over the thermal bath and 
where the operators depend on time $t$ according to an Heisenberg 
representation with respect to the free Hamiltonian $H_{0}=H_{v}+H_{b}$. 
At this step, the relaxation rate can be written in an improved form by 
inserting the intermediate basis set (Eq.(\ref{eq:bloch1})) for which 
the matrix elements of the population operator are defined as 
\begin{equation}
\langle k,m \mid b^{+}_{n}b_{n} \mid k',m' \rangle=\frac{2}{N}\delta_{mm'} e^{i(k-k')n}cos(\frac{k-k'}{2}m)
\label{eq:pop}
\end{equation}
Therefore, by inserting Eq.(\ref{eq:pop}) into Eq.(\ref{eq:rate2}), the relaxation rate is written as   
\begin{eqnarray}
& &W^{(2)}_{k,B} = \frac{8}{N^{2}}Re \sum_{n,n'} \sum_{m,m'} 
\sum_{k'} \int_{0}^{\infty} dt e^{i\omega_{B}(k)t}
 \nonumber \\
& &\langle \Delta \omega_{n}(t) \Delta \omega_{n'}(0) \rangle e^{i(k-k')(n-n')} \langle k',m \mid e^{-iH_{v}t} \mid k',m' \rangle  \nonumber \\ 
& & \Psi_{k}^{B*}(m) \Psi_{k}^{B}(m') cos((k-k')m/2)cos((k-k')m'/2) 
\label{eq:rate3}
\end{eqnarray}

The TVBS relaxation rate depends on the correlation functions of the coupling $\Delta \omega $. The characteristic time of this rate is the correlation time $\tau_{c} $ of the thermal bath which corresponds to the time for which the correlation functions vanish. In a general way, $\tau_{c}$ is about 1 ps for phonons in low-dimensional molecular lattices. For instance, molecular dynamics simulations carried out for the system CO/NaCl found a correlation time value equal to 0.6 ps at 25 K \cite{kn:pouthier3}. We thus assume that this time scale is small compared to the characteristic time evolution of the vibrons \cite{kn:pouthier2}. Indeed, this later time is of about the invert of the hopping constant $\Phi$ which is generally lesser than or about to 10 cm$^{-1}$ for both molecular adsorbates and proteins. For instance, the dispersion of the stretching vibration Si-H of the system Si/H(111) was found to be about 9 cm$^{-1}$ \cite{kn:sih}. In the case of the $(2 \times 1)$ CO monolayer adsorbed on NaCl, the lateral interaction was shown to be strongly anisotropic with a vibron bandwidth equal to 3 cm$^{-1}$ along the longer size of the unit cell and to 12 cm$^{-1}$ along the perpendicular direction \cite{kn:pouthier3}. In a similar way, for $\alpha$-helix proteins, the accepted value for the vibron hopping constant is equal to 7.8 cm$^{-1}$ (see for instance Refs. \cite{kn:brizhik,kn:pang}).

Consequently, the inequality $\Phi \times \tau_{c} << 1 $ allows us two simplifying approximations in Eq.(\ref{eq:rate3}). First, we assume that the thermal bath does not exhibit spatial correlations ($\langle \Delta \omega_{n}(t) \Delta \omega_{n'}(0) \rangle =
\langle \Delta \omega_{n}(t) \Delta \omega_{n}(0) \rangle \delta_{nn'}$) and second, we neglect the non-diagonal elements of the free propagator $exp(-iH_{v}t)$. The second assumption means that over a time scale of about $\tau_{c} $, a vibron does not have enough time to make a transition to a neighboring site. 

As a result, by inserting Eqs.(\ref{eq:Deltaom}) and (\ref{eq:J1}) into Eq.(\ref{eq:rate3}) and by performing both the time integration and the thermal average, the TVBS relaxation rate is finally expressed as 
\begin{eqnarray}
W^{(2)}_{k,B}&=& 8\pi \int d\Omega J(\Omega)[n(\Omega)\rho^{B}_{k}(\omega_{B}(k)+\Omega) \nonumber \\ 
& &+(n(\Omega)+1) \rho^{B}_{k}(\omega_{B}(k)-\Omega)]
\label{eq:rate4}
\end{eqnarray}
where $n(\Omega)=(exp(-\Omega/kT)-1)^{-1}$ denotes the Bose-Einstein phonon distribution at temperature $T$ and where $\rho^{B}_{k}(\omega)$ is the projected two-vibron density of states defined in terms of the two-vibron Green operator $G(\omega)=(\omega-H_{v})^{-1}$ as
\begin{eqnarray}
\rho^{B}_{k}(\omega)&=&-Im \frac{1}{\pi N} \sum_{m}\sum_{k'} 
\langle k',m \mid G(\omega) \mid k',m \rangle \nonumber \\
& &\mid \Psi^{B}_{k}(m) cos((k-k')m/2) \mid^{2}
\label{eq:DOS} 
\end{eqnarray}
The projected density of states (DOS) $\rho^{B}_{k}(\omega)$ represents the coupling distribution between the TVBS with wave vector $k$ and energy $\omega_{B}(k)$ and all the other two-vibron states which the energy ranges between $\omega$ and $\omega +d \omega$. Such a coupling is mediated by the interaction between the vibrons and the phonons of the thermal bath. 

Note that the projected DOS can be expressed as the sum of two contributions, i.e. $\rho^{B}_{k}(\omega)=\rho^{B \rightarrow B}_{k}(\omega)+\rho^{B \rightarrow F}_{k}(\omega)$, where $\rho^{B \rightarrow B}_{k}(\omega)$ and $\rho^{B \rightarrow F}_{k}(\omega)$ characterize the coupling distribution with the TVBS and the TVFS, respectively. The projected DOS $\rho^{B \rightarrow B}_{k}(\omega)$ can be easily determined by considering the restriction of the two-vibron Green operator to the TVBS in Eq.(\ref{eq:DOS}). It is expressed as  
\begin{eqnarray}
\rho^{B \rightarrow B}_{k}(\omega)&=&-Im \frac{1}{\pi N} \sum_{m}\sum_{k'} 
\frac{\mid \Psi^{B}_{k'}(m) \mid^{2}}{\omega-\omega_{B}(k')+i0^{+}} \nonumber \\
& &\mid \Psi^{B}_{k}(m) cos((k-k')m/2) \mid^{2}
\label{eq:DOS1} 
\end{eqnarray}
Note that the knowledge of both $\rho^{B }_{k}(\omega)$ and $\rho^{B \rightarrow B}_{k}(\omega)$ allows us to calculate $\rho^{B \rightarrow F}_{k}(\omega)$. As a result, since the TVBS relaxation rate depends linearly on the projected DOS, it can be expressed as the sum of two contributions as
\begin{equation}
W^{(2)}_{k,B}=W^{(2)}_{k,B}(B \rightarrow B)+W^{(2)}_{k,B}(B \rightarrow F)
\label{eq:RATEFIN} 
\end{equation}
where $W^{(2)}_{k,B}(B \rightarrow B)$ and $W^{(2)}_{k,B}(B \rightarrow F)$ characterize the relaxation of the TVBS with wave vector $k$ and energy $\omega_{B}(k)$ over all the other bound and free states, respectively (Fig. 2b). 

In order to compare the single and two-vibron relaxation rates, let us mention that the rate for a single-vibron state with wave vector $q$ and frequency $\omega(q)$ can be straightforwardly determined by using the previous procedure. Indeed, after performing some algebraic manipulations, we thus obtain
\begin{eqnarray}
W^{(1)}_{q}& =& 2\pi \int d\Omega J(\Omega) [n(\Omega)g(\omega(q)+\Omega) \nonumber \\ 
& &+(n(\Omega)+1) g(\omega(q)-\Omega)]
\label{eq:rateSV}
\end{eqnarray}
where $g(\omega)$ stands for the single-vibron density of states defined as
\begin{equation}
g(\omega)=\frac{1}{2\pi\Phi}\frac{1}{\sqrt{1-(\frac{\omega-\omega_{0}}{2\Phi})^{2}}} 
\label{eq:g1v}
\end{equation}

As shown in Eqs.(\ref{eq:rate4}) and (\ref{eq:rateSV}), the rates depend on the temperature via their dependence with respect to the average number of phonons. Moreover, both the two-vibron and single-vibron rates exhibit two contributions connected to the absorption (term proportional to $n(\Omega)$) and to the emission (term proportional to $n(\Omega)+1$) of an acoustic phonon, respectively. As pointed out by Jakob and Persson \cite{kn:jakob3}, a factor 4 discriminates between the two rates. However, Eqs.(\ref{eq:rate4}) and (\ref{eq:rateSV}) show that the main difference between the dephasing mechanism related to single and two-vibron states appears through the dependence of the rates with respect to the single and two-vibron density of states. Although the single-vibron density of states can be calculated exactly (Eq.(\ref{eq:g1v})), the projected DOS $\rho^{B}_{k}(\omega)$ still remains unknown. Therefore, the next section is devoted to its calculation based on the determination of the two-vibron Green operator \cite{kn:pouthier1}. 

\subsection{Green function calculation} 

Within the intermediate basis $\mid k,m \rangle$, the Green operator $G(\omega)=(\omega-H_{v})^{-1}$ is block diagonal and can be evaluated for each wave vector $k$. Let us define $G_{k}$ as the two-vibron Green operator connected to the restriction $H_{vk}$ of $H_{v}$ to the subspace $\{\mid k,m \rangle\}$, $m=0,...,(N-1)/2$. It corresponds to the Green function associated to the Schrodinger equation Eqs.(\ref{eq:schrod1})-(\ref{eq:schrod3}). Therefore, $G_{k}$ can be determined explicitely by taking advantage of the equivalence between the Schrodinger equation Eqs.(\ref{eq:schrod1})-(\ref{eq:schrod3}) and the dynamics of a single particle \cite{kn:pouthier1}. Such a determination can be performed by applying the procedure introduced by L. Dobrzynski \cite{kn:dob} to calculate the response function of super-lattices and composite materials.

To proceed, we first consider the tight-binding problem on the lattice shown in Fig. 1b which corresponds to an infinite chain. This chain is denoted the ideal lattice by 
opposition to the real lattice drawn in Fig. 1a. Let $H^{\infty}_{k}$ denotes the tight-binding Hamiltonian which characterizes the dynamics of this ideal lattice. The associated Green function, $G^{\infty}_{k}(\omega)=(\omega-H^{\infty}_{k})^{-1}$, can be calculated analytically as
\begin{equation}
G^{\infty}_{k}(m,m')=-\frac{(a \pm \sqrt{a^{2}-1})^{\mid m-m' \mid}}
{\pm 2\tilde{\Phi}_{k} \sqrt{a^{2}-1}}
\label{eq:G0}
\end{equation}
where $a=(\omega-2\omega_{0})/\tilde{\Phi}_{k}$. In Eq.(\ref{eq:G0}) the sign $\pm$ is chosen in order to ensure the regular behavior of the Green function when $\mid m-m' \mid \rightarrow \infty$.
The Hamiltonian $H_{vk}$ can be obtained by adding an operator $V_{k}$ to $H^{\infty}_{k}$ and by taking the restriction to the sites which belong to the real lattice as
\begin{equation}
H_{vk}=P_{k}(H^{\infty}_{k}+V_{k})P_{k}
\label{eq:green1}
\end{equation}
where $P_{k}=\sum_{m=0}^{(N-1)/2} \mid k,m \rangle \langle k,m \mid$ is the projector onto the real lattice. The goal of the operator $V_{k}$, known as the cleavage operator \cite{kn:dob}, is first to disconnect in the ideal lattice the sites which do not belong to the real lattice, and then to introduce the real boundary conditions. 
The cleavage operator acts in the subspace connected to the interface between the real and the ideal lattice, i.e. $m=-1,0,1$. Inside this subspace, it is straightforward to show that the non vanishing elements of the cleavage operator are expressed as  
\begin{equation}
V_{k}= \left(
\begin{array}{ccc}
0 & -\tilde{\Phi}_{k} & 0  \\
-\tilde{\Phi}_{k} & -2A & (\sqrt{2}-1)\tilde{\Phi}_{k} \\
0 & (\sqrt{2}-1)\tilde{\Phi}_{k} & 0 
\end{array}
\right)  
\label{eq:cleavage}
\end{equation}
Finally, from Eq.(\ref{eq:green1}), the Green function of the real lattice is thus expressed in terms of the Green function $G^{\infty}_{k}(\omega)$ of the ideal lattice, as
\begin{equation}
G_{k}=P_{k}[G^{\infty}_{k}+G^{\infty}_{k}V_{k}P_{k}(1-G^{\infty}_{k}V_{k})^{-1}P_{k}G^{\infty}_{k}]P_{k}
\label{eq:green2}
\end{equation}
Therefore, by inserting Eqs. (\ref{eq:G0}) and (\ref{eq:cleavage}) into Eq.(\ref{eq:green2}),
the two-vibron Green operator can be computed exactly.

At this step, the knowledge of the two-vibron Green function allows us to determine the two-vibron projected DOS (Eq.(\ref{eq:DOS})) and then to compute the TVBS relaxation rate (Eq.(\ref{eq:rate4})). Such a calculation is illustrated in the following section. 
 
\section{Numerical results}

In this section, the previous formalism is applied to characterize the TVBS relaxation rate 
for a one-dimensional molecular lattice. The lateral force constant $\Phi$, fixed to 1.0, will be used as the frequency unit. Since the vibron dispersion in low-dimensional molecular lattices is rather small (lesser than or of about 10 cm$^{-1}$), we assume that it can be neglected 
with respect to the phonon bandwidth, i.e. $\Omega_{c}>>\Phi$. Moreover, for small molecules, the anharmonic parameter is usually close to the gas phase value and ranges between $10-30$ cm$^{-1}$. We thus consider that a reasonable value for the maximum of the reduced parameter $A/\Phi$ is of about 10. Finally, in order to compare the single and two-vibron relaxation rates, the single-vibron state with zero wave vector ($q=0$) and with frequency $\omega(q=0)=\omega_{0}+2\Phi$ will be used as the reference for the calculation of  $W^{(1)}_{q=0}$ (Eq.(\ref{eq:rateSV})). 
 
The behavior of the projected DOS $\rho^{B}_{k}(\omega)$ as a function of the anharmonic parameter $A$ is shown in Fig. 3 for two values of the wave vector $k=0$ (full line) 
and $k=\pi/2$ (dotted line). The projected DOS extends from 
$2\omega_{0}-2\sqrt{A^{2}+4\Phi^{2}}$ to $2\omega_{0}+4\Phi$. It is maximum in the 
frequency range of the bound states, i.e. $2\omega_{0}-2\sqrt{A^{2}+4\Phi^{2}} < \omega < 2\omega_{0}-2A$, and exhibits two peaks located at the sides of the TVBS band. These peaks characterize the Van Hove singularities which occur when the group velocity of the TVBS vanishes.  
When $A<2$, the TVBS band and the TVFS band overlap each other. As a result, the 
projected DOS forms a single band which is significant over the entire frequency 
range. By contrast, when $A>2$, the projected DOS exhibits two bands located 
in the frequency range of the TVBS and TVFS, respectively. As increasing the anharmonicity, 
the weight of the projected DOS in the frequency range of the TVFS decreases drastically. Note that the projected DOS becomes almost independent on the value of the wave vector $k$ when $A>>2$. 

The behavior of the TVBS relaxation rate $W^{(2)}_{k=0,B}$ as a function of the temperature is shown in Fig. 4a for different values of the anharmonicity $A=1,3,5,7$ and for a bath dimension $\alpha=1$. At very low temperature, i.e. $kT << \Phi$, the rate exhibits a powerlaw dependence with respect to the temperature. By contrast, as increasing the temperature, the rate behaves linearly versus $T$ with a slope which increases as the anharmonicity $A$ increases. In Fig. 4b, the temperature dependence of the ratio $W^{(2)}_{k=0,B}/W^{(1)}_{q=0}$ is shown for the same set of values of the anharmonicity. The various curves exhibit two distinct regimes depending on the temperature. At low temperature, the ratio increases in a quasi-linear way with a slope which increases with $A$. Then, at higher temperature, the ratio becomes almost temperature independent and reaches a limiting value which strongly depends on the anharmonic parameter. It increases as the anharmonicity increases and evolves from $2.79$ when $A=1.0$ to $3.85$ when $A=7.0$. Note that the transition between the two previous regimes is independent on the anharmonicity and occurs when $kT/\Phi$ is of about 15-20. 
 
As shown in Figs. 5a and 5b, both the relaxation rate $W^{(2)}_{k=0,B}$ and the ratio
$W^{(2)}_{k=0,B}/W^{(1)}_{q=0}$ exhibit a similar temperature dependence when the 
dimension of the thermal bath is set to $\alpha=2$. The transition between the two regimes occurs at almost the same temperature, i.e. $kT/\Phi \approx 15-20$ (Fig. 5b). However, as shown in Fig. 5a, the linear evolution of the rate with respect to the temperature is characterized by a slope which decreases as the anharmonicity increases, in marked contrast with the previous situation. In addition, the limiting values reached by the ratio $W^{(2)}_{k=0,B}/W^{(1)}_{q=0}$ at high temperature (Fig. 5b) decreases as $A$ increases and evolves from $4.42$ when $A=1.0$ to $1.56$ when $A=7.0$. Note that both the relaxation rate $W^{(2)}_{k=0,B}$ and the ratio $W^{(2)}_{k=0,B}/W^{(1)}_{q=0}$ exhibit a similar behavior when the bath dimension is set to $\alpha=3$ (not drawn).   

When the bath dimension is equal to $\alpha=1$, the behavior of the relaxation rate with respect to the anharmonic parameter $A$ is shown in Fig. 6a for $kT/\Phi=1$ (circle), 10 (square), 50 (triangle), 100 (diamond) and for $k=0$ (full line) and $k=\pi/2$ (dotted line). Whatever both the temperature and the value of the wave vector, the ratio $W^{(2)}_{k,B}/W^{(1)}_{q=0}$ increases with the anharmonicity and reaches a limiting value for a strong anharmonicity. At low temperature, i.e. $kT=\Phi$, this limiting value is almost equal to unity whereas it appears close to 4 at high temperature. Note that the rate for the TVBS with wave vector $k=0$ is lesser than the rate for the TVBS with wave vector $k=\pi/2$. Nevertheless, the difference is less and less important as the anharmonicity becomes stronger. 
In Fig. 6b, the behavior of the ratio $W^{(2)}_{k,B}(B\rightarrow F)/W^{(2)}_{k,B}(B\rightarrow B)$ with respect to $A$ is shown for the same set of temperatures. For both $k=0$ and $k=\pi/2$, the ratio decreases as the anharmonicity increases and increases with the temperature. It is almost lesser than unity excepted at high temperature and low anharmonicity. Note that when $kT/\Phi>10$, the curves decrease with respect to $A$ according to an invert powerlaw. For instance, when $kT=50\Phi$ and $k=0$, the ratio evolves from $1.02$ when $A=1$ to $0.014$ when $A=10$.
     
As shown in Figs. 7a and 7b, the relaxation rate behaves in a fully different way when the bath dimension is set to $\alpha=2$. Indeed, although the ratio $W^{(2)}_{k,B}/W^{(1)}_{q=0}$ appears almost independent on the anharmonicity at very low temperature ($kT=\Phi$), it decreases as the anharmonic parameter $A$ increases when $kT/\Phi=10,50,100$ (Fig. 7a), in a marked contrast with the previous results. For instance, when $kT=50\Phi$ and $k=0$, the TVBS relaxation rate is greater than the single-vibron relaxation rate by a factor 4.17 when $A=1$ whereas this ratio decreases to 0.98 when $A=10$. In addition, the relaxation rate for the TVBS with wave vector $k=0$ appears now greater than the relaxation rate for the TVBS with wave vector $k=\pi/2$ but, as previously, the difference is less and less important as the anharmonicity increases. 
As shown in Fig. 7b, the ratio $W^{(2)}_{k,B}(B\rightarrow F)/W^{(2)}_{k,B}(B\rightarrow B)$ decreases as the anharmonicity increases. However, it is greater than unity over a large  anharmonicity range, excepted at low temperature. For instance, when $kT=10\Phi$ and $k=0$, the ratio, which is equal to $3.31$ when $A=1$, becomes lesser than unity when $A>4.0$. Note that, the various curves exhibit two regimes depending on the anharmonicity since the ratio rapidly decreases at low anharmonicity, i.e. $A<2.5$, whereas it decreases more slowly at high anharmonicity.

The same results are shown in Figs. 8a and 8b when the bath dimension is equal to $\alpha=3$. As in the previous case, the ratio $W^{(2)}_{k,B}/W^{(1)}_{q=0}$ is almost independent on the anharmonicity at very low temperature ($kT=\Phi$), but decreases as the anharmonicity increases when $kT/\Phi=10,50,100$ (Fig. 8a). For instance, when $kT=50\Phi$ and $k=0$, the TVBS relaxation rate is equal to $7.01$ time the single-vibron rate when $A=1$ whereas this ratio is equal to $3.89$ when $A=10$. Moreover,  
the relaxation rate for the TVBS with wave vector $k=0$ is greater than the relaxation rate for the TVBS with wave vector $k=\pi/2$. Note that, at high temperature, the rate clearly shows two distinct regimes since it exhibits a fast decrease at low anharmonicity ($A<3.5$) whereas it decreases more slowly at high anharmonicity and becomes almost constant.    
As shown in Fig. 8b, the ratio $W^{(2)}_{k,B}(B\rightarrow F)/W^{(2)}_{k,B}(B\rightarrow B)$ behaves in a fully different way with respect to the two previous situations. At low temperature, i.e. $kT=\Phi$, it is lesser than 
unity and decreases as the anharmonicity increases. When the temperature is equal to $kT=10\Phi$, it first slightly decreases at very low anharmonicity ($A<2$), then it increases with $A$ to reach a maximum value equal to $10.45$ at $A \approx 7.0$. Finally, as increasing the anharmonicity from the previous value, the ratio decreases. At higher temperature, i.e. $kT/\Phi=50$ and $100$, the ratio shows a similar behavior but increases more drastically. Note that it does not reach a maximum value into the anharmonicity range considered in this work.

\section{Discussion}

To interpret and discuss the previous numerical results, let us first focus our attention on the temperature dependence of the TVBS relaxation rate. Indeed, 
whatever both the anharmonicity and the bath dimension, the rate exhibits a powerlaw and a linear dependence on the temperature at low and 
high temperature, respectively. These features are due to the fact that the relaxation involves single phonon processes in the course of which a phonon is absorbed or emitted. 
As a result, the rate depends linearly on the phonon Bose-Einstein distribution 
which is responsible for the observed behavior. When $k=0$, the TVBS frequency 
lies at the bottom of the TVBS band. The relaxation involves phonon absorption only, and, therefore, the corresponding rate vanishes as 
the temperature tends to zero since the number of phonons becomes infinitesimally 
small. By contrast, for a finite wave vector, phonon emission can occur leading 
to a finite value of the rate at zero temperature. 
Note that in our simulation we used a positive value for the lateral force constant $\Phi$. As a 
result, the single vibron state $q=0$, which was considered as the reference for 
the calculation of the rate $W^{(1)}_{q=0}$, lies at the top of the single-vibron band. Therefore, $W^{(1)}_{q=0}$ reaches a finite value at $T=0$ since phonon emission 
can takes place and depends linearly on the temperature at high temperature. 

The more surprising results are clearly related to the behavior of the TVBS relaxation rate with respect to the anharmonicity. Indeed, whatever the value of the bath dimension, the rate is a slowly varying function of the anharmonicity at very low temperature, and the relaxation path involves the decay of the TVBS over the other TVBS, only (See Figs. 7b, 8b and 9b, for $kT=\Phi$). However, at sufficiently high temperature ($kT > 10 \Phi$), the numerical results indicate that the rate exhibits two distinct regimes depending on the bath dimension. When $\alpha=1$, the TVBS relaxation rate increases with the anharmonicity and reaches a limiting value almost four time greater than the single-vibron rate. Moreover, the main contribution of the rate involves the relaxation over the other TVBS, only. In a marked contrast, when the bath dimension is equal to 2 and 3, the rate decreases as the anharmonic parameter $A$ increases. When $\alpha=2$, the relaxation involves the decay of the TVBS into both TVBS and TVFS, the two processes contributing in a similar way to the rate. By contrast, when $\alpha=3$, the decay into the TVBS becomes negligible and the main contribution of the rate involves transitions from the TVBS to the continuum of TVFS.

Such features originate in the convolution product between the projected DOS $\rho_{k}^{B}(\omega)$ and the coupling distribution $J(\omega)$ weighted by the Bose-Einstein distribution. When $kT << \Phi$, the Bose-Einstein distribution decreases exponentially with the phonon frequency and thus favors the exchange of very low frequency phonons. Therefore, whatever the bath dimension, the TVBS cannot decay into the TVFS continuum and relaxes over the other TVBS. When $kT >> \Phi$, both phonon emission and phonon absorption contribute similarly to the rate since $n(\Omega) \approx n(\Omega)+1 \approx kT/\Omega$. The rate reduces to the convolution between the projected DOS and the distribution $kTJ(\Omega)/\Omega$. This latter function scales as $\Omega^{\alpha-1}$ and is constant when $\alpha = 1$ whereas it increases with $\Omega$ when both $\alpha=2$ and $3$. 

Therefore, when $\alpha=1$, the relaxation of the TVBS mediated by phonon exchange occurs with the same probability whatever the value of the frequency of the phonon involved in the process. As a consequence, the behavior of the rate is governed by the strength of the coupling between the TVBS and the other two-vibron states induced by the vibron/phonon interaction. From Eqs.(\ref{eq:rate4})-(\ref{eq:DOS}), this coupling, which is expressed as the integration of 
the projected DOS, leads to a relaxation rate written as 
\begin{equation}
W^{(2)}_{k,B}= \frac{2\pi kT J_{0}}{N \Omega_{c}}
\sum_{m,k'} \mid \Psi^{B}_{k}(m) cos((k-k')m/2) \mid^{2}
\label{eq:rate1D}
\end{equation}
After some straightforward calculations, Eq.(\ref{eq:rate1D}) leads to a ratio  $W^{(2)}_{k,B}/W^{(1)}_{q=0}$ expressed as
\begin{equation}
\frac{W^{(2)}_{k,B}}{W^{(1)}_{q=0}}=2+\frac{2A}{\sqrt{ A^{2}+(2\Phi cos(k/2))^{2}}}
\label{eq:ratio}
\end{equation}
The ratio $W^{(2)}_{k,B}/W^{(1)}_{q=0}$ increases with the anharmonicity, in perfect agreement with the numerical results shown in Fig. 6a , and evolves from 2 when $A=0$ to 4 when A tends to infinity. In other words, for a small anharmonicity, the TVBS relaxation rate is twice the single-vibron rate which indicates that the two vibrons are almost independent, i.e. the TVBS rate appears as the sum of the rates connected to each vibron. As increasing the anharmonic parameter, the length of the bond between the two vibrons (Eq. (\ref{eq:xi})) reduces and the local nature of the vibron population increases to reach a limiting value equal to 2. The strength of the coupling, proportional to the square of the population, leads to a factor 4 between the TVBS rate and the single-vibron rate. Note that as increasing the value of the wave vector, the local nature of the TVBS increases since the length of the bond between the two vibrons decreases (cf Eq.(\ref{eq:xi})) leading to an enhancement of the relaxation rate. 

In a marked contrast, when $\alpha=2$ and $3$, the behavior of the TVBS relaxation rate results in the competition between the projected DOS and the coupling distribution $kTJ(\Omega)/\Omega$ which contribute in an opposite way. 
Indeed, the projected DOS, which is maximum in the frequency range of the bound states, favors transitions from the TVBS to the other TVBS. Such transitions can thus be induced by the exchange of low frequency phonons. However, the coupling distribution $kTJ(\Omega)/\Omega$, which increases with the phonon frequency, favors transitions mediated by high frequency phonons from the TVBS to the TVFS continuum. 
At high temperature, the numerical results clearly show that the decay of the TVBS into the TVFS continuum is the main contribution of the rate. This feature indicates that the influence of the phonon coupling distribution prevails. However, the contribution  of the projected DOS in the frequency range of the TVFS decreases drastically as the anharmonicity increases and counterbalances the growth of the phonon coupling distribution. As a result, the rate decreases with respect to the anharmonic parameter. The effect of the phonon coupling distribution becomes stronger as the bath dimension increases. Indeed, when $\alpha=2$, both decays into the TVBS and TVFS continuum, occur with almost the same probability. However, when $\alpha=3$, the latter path for the relaxation becomes strongly dominant. Note that as increasing the value of the wave vector, the energy difference between the TVBS and the TVFS continuum decreases. As a result, the frequency of the phonon exchanged during the relaxation mechanism decreases leading to a lesser value for the relaxation rate since the coupling distribution $J(\Omega)$ is reduced.
 
When the anharmonicity is sufficiently strong, an analytic expression of the projected DOS in the frequency range of the TVBS can be determined. 
Indeed, by performing a second order expansion of Eq.(\ref{eq:TVBSE}) with respect to
$\Phi /A$, the frequency of the TVBS is expressed as $\omega_{B}(k) \approx 2\omega_{0}-2A-2(\Phi^{2}/A)(1+cos(k))$. Therefore, the TVBS dynamics can be compared with the dynamics of a single vibron characterized by a force constant $\Phi' = -\Phi^{2}/A$. The projected DOS
$\rho^{B \rightarrow B}(\omega)$ is thus given by the single vibron density of states Eq.(\ref{eq:g1v}) according to the correspondences $\Phi \rightarrow \Phi^{2}/A$ and $\omega_{0} \rightarrow 2\omega_{0}-2A-2\Phi^{2}/A$. The rate $W^{(2)}_{k,B}(B\rightarrow B)$ for the decay of the TVBS into the other TVBS is thus given by Eq.(\ref{eq:rateSV}) with the previous correspondences and by adding a factor four to account for the population effect. For instance, at high temperature, the rate for the relaxation of the TVBS with zero wave vector is written 
as  
\begin{equation}
W^{(2)}_{k=0,B}(B\rightarrow B)=\frac{8J_{0}kT}{\Omega_{c}}
(\frac{2\Phi^{2}}{A\Omega_{c}})^{\alpha-1} I(\alpha-1)
\label{eq:RBB}
\end{equation}
where $I(\alpha)$ is defined in terms of the Gamma function as $I(\alpha)=2^{3\alpha}\Gamma(\alpha+1/2)^{2}/\Gamma(2\alpha+1)$.
As shown in Eq.(\ref{eq:RBB}), the rate $W^{(2)}_{k=0,B}(B\rightarrow B)$, which scales as $A^{1-\alpha}$, is constant when $\alpha=1$ whereas it decreases when $\alpha>1$. Although no analytical expression for the rate $W^{(2)}_{k,B}(B\rightarrow F)$ can be determined, the numerical results clearly show that this rate scales as 1/A when $\alpha=1$ and becomes negligible compared with $W^{(2)}_{k,B}(B\rightarrow B)$. By contrast, when $\alpha=2$ and $3$, the rate $W^{(2)}_{k,B}(B\rightarrow F)$ decreases with $A$ according to a law decaying more slowly than $A^{1-\alpha}$, and is the dominant contribution of the rate. 

At this step, let us mention that the numerical results presented in the previous section were obtained by considering integer values for the bath dimension. However, the formalism can be applied to real values of the bath dimension. In particular, at high temperature and strong anharmonicity, our calculations predict that the TVBS relaxation rate decreases with the anharmonicity when $\alpha>1$. In addition, the dominant relaxation pathway corresponds to the decay of the TVBS into the other TVBS over a large $\alpha$ range, i.e. $1<\alpha <\alpha_{c}$ where $\alpha_{c}<2$. In that case, the total relaxation rate is thus approximately given by Eq.(\ref{eq:RBB}). For instance, when $kT=100\Phi$ and $A=10$, the ratio $W^{(2)}_{k,B}(B\rightarrow B)/W^{(2)}_{k,B}(B\rightarrow F)$ is equal to unity for a bath dimension $\alpha_{c} \approx 1.9$.

To illustrate the previous theoretical calculations, let us compare our results with experimental data. The system CO/NaCl(100) was previously studied using molecular dynamics simulations \cite{kn:pouthier3}. The CO monolayer has a $(2 \times 1)$ structure with a rectangular unit cell containing two molecules mutually perpendicular. Since the lateral interaction was shown to be strongly anisotropic, the vibron dynamics of the monolayer exhibits a one-dimensional character. For instance, along the longer size of the unit cell, the lateral force constant between nearest neighbor molecules is positive and equal to $\Phi=2.94$ cm$^{-1}$. Assuming an anharmonic parameter equal to the gas phase value, i.e. $A=13.3$ cm$^{-1}$, the reduced anharmonicity is $A/\Phi=4.5$. At $T=5$ K and for a bath dimension $\alpha=1$, the theoretical ratio $W^{(2)}_{k=0,B}/W^{(1)}_{q=0}$ is equal to 1.3. Note that for a bath dimension equal to 2 or 3, the ratio is drastically reduced by a factor of about 10.   
This result is in a perfect agreement with the experimental ratio, equal to 1.55, obtained by Dai and Ewing \cite{kn:dai}. A genuine one-dimensional lattice was obtained by Jakob \cite{kn:jakob2} 
by carefully controlling the adsorption of CO on a Pt(111) surface exhibiting steps. As measured by the author, the dynamical coupling between the molecules is responsible for a blueshift $\Delta\omega=19$ cm$^{-1}$. By assuming nearest neighbor interactions, this coupling leads to a positive lateral force constant $\Phi=9.5$ cm$^{-1}$. In addition, the anharmonicity of the system, measured by Jakob, yields a reduced parameter $A/\Phi=1.42$. By following the experiment performed by the author, we thus obtain a ratio $W^{(2)}_{k=0,B}/W^{(1)}_{q=0}=2$ at $T=80$ K, which is in good agreement with the experimental ratio equal to 1.5.

To conclude, let us discuss the main implications of the previous results on the energy transport mediated by TVBS. Indeed, from a quantum kinetic point of view, the vibron/phonon interaction is responsible for the breaking of the coherence and favors a diffusional motion of the vibrons after a time scale of about the dephasing time \cite{kn:pouthier2}. The previous results clearly show that such a diffusional motion, for TVBS, drastically depends on the bath dimension. When $\alpha=1$, the coupling with the phonons is not strong enough to break the bond between the two vibrons which cannot decay into the continuum of TVFS. Therefore, the thermal bath keeps the internal coherence between the two vibrons but induces momentum scattering which is responsible for transitions from a TVBS to another TVBS. Such a scattering in momentum space corresponds to incoherent hops in real space in the course of which the center of mass of the two vibrons has a diffusional motion. Therefore, although the energy is transported in an incoherent way, it diffuses by packets along the lattice. Note that since the TVBS relaxation rate is almost four time greater than the corresponding single-vibron rate, a similar ratio is expected for the diffusion coefficients. By contrast, when $\alpha=2$ or $3$, the phonon bath favors the breaking of the bond between the two vibrons over a large anharmonicity range. As a result, the decay of the TVBS into the continuum of TVFS takes place and the two particles will tend to diffuse independently leading to the uniformization of the energy along the lattice. 

Finally, let us mention some remarks on forthcoming work. First, the present results were obtained by considering the coupling between the vibrons and acoustic phonons. However, as discussed in sec. II, the formalism can be applied to different kinds of thermal bath. In particular, it would be interesting to investigate the effect of optical phonons which can be responsible for the occurrence of resonances between the TVBS band and the TVFS continuum. These features were pointed out experimentally by Jakob et al. \cite{kn:jakob3} when studying the linewidth of the overtone of the system CO/Ru and attributed to the coupling with the frustrated translation mode of the monolayer. Second, we have restricted our analysis to processes involving a single phonon, only. However, two-phonon processes can contribute significantly to the relaxation rate, depending on the bath dimension. For instance, two-phonon processes may enhance the spectral distribution of the vibron/phonon coupling at low frequency favoring the 
decay of TVBS into the TVBS band. 

\section{ACKNOWLEDGMENTS}

The author would like to thank Professor C. Girardet for fruitful discussions
and for a critical reading of the text.

\newpage 

\begin{center}
\textbf{Figure Caption}
\end{center}

Figure 1 : Equivalence between the Schrodinger equation of two-vibron states in the intermediate basis and the dynamics of a single fictitious particle moving quantum mechanically in a one-dimensional lattice (see the text). The index $m$, which characterize the separating distance between the two vibrons, labels the sites of the equivalent lattice. (a) The equivalent lattice, called the real lattice, extends from $m=0$ to $m=(N-1)/2$. (b) Ideal lattice used to calculate the two-vibron Green operator. The circles define the self-frequencies of the sites $m$ and the lines characterize the hopping constants $\tilde{\Phi_{k}}$ (thin line) and $\sqrt{2}\tilde{\Phi_{k}}$ (full line) (see the text).

Figure 2 : (a) Two-vibron eigenstates energy spectrum which exhibits two bands. The high frequency band, located around $2\omega_{0}$, contains the two-vibron free states continuum. The low frequency band is the dispersion relation of the two-vibron bound states. The anharmonic parameter is equal to $A=2.5\Phi$. (b) Schematic representation of the relaxation processes.  

Figure 3 : Projected DOS $\rho^{B}_{k}(\omega)$ as a function of the anharmonic parameter $A$ for two values of the wave vector $k=0$ (full line) and $k=\pi/2$ (dotted line). 

Figure 4 : (a) Temperature dependence of the two-vibron bound state relaxation rate $W^{(2)}_{k=0,B}$ for different values of the anharmonicity $A=1$ (circle), 3 (square), 5 (trinagle), 7 (diamond) and for a bath dimension $\alpha=1$. (b) Temperature dependence of the ratio $W^{(2)}_{k=0,B}/W^{(1)}_{q=0}$ for different values of the anharmonicity $A=1$ (circle), 3 (square), 5 (trinagle), 7 (diamond) and for a bath dimension $\alpha=1$. 

Figure 5 : (a) Temperature dependence of the two-vibron bound state relaxation rate $W^{(2)}_{k=0,B}$ for different values of the anharmonicity $A=1$ (circle), 3 (square), 5 (trinagle), 7 (diamond) and for a bath dimension $\alpha=2$. (b) Temperature dependence of the ratio $W^{(2)}_{k=0,B}/W^{(1)}_{q=0}$ for different values of the anharmonicity $A=1$ (circle), 3 (square), 5 (trinagle), 7 (diamond) and for a bath dimension $\alpha=2$. 
. 

Figure 6 : (a) Behavior of the TVBS relaxation rate with respect to the anharmonicity $A$ for $kT/\Phi=1$ (circle), 10 (square), 50 (triangle), 100 (diamond) and for $k=0$ (full line) and $k=\pi/2$ (dotted line). The bath dimension is equal to $\alpha=1$. (b) Behavior of the ratio $W^{(2)}_{k,B}(B\rightarrow F)/W^{(2)}_{k,B}(B\rightarrow B)$ with respect to the anharmonicity $A$ for the same set of parameters as in Fig 6a.  

Figure 7 : (a) Behavior of the TVBS relaxation rate with respect to the anharmonicity $A$ for $kT/\Phi=1$ (circle), 10 (square), 50 (triangle), 100 (diamond) and for $k=0$ (full line) and $k=\pi/2$ (dotted line). The bath dimension is equal to $\alpha=2$. (b) Behavior of the ratio $W^{(2)}_{k,B}(B\rightarrow F)/W^{(2)}_{k,B}(B\rightarrow B)$ with respect to the anharmonicity $A$ for the same set of parameters as in Fig 7a. 

Figure 8 : (a) Behavior of the TVBS relaxation rate with respect to the anharmonicity $A$ for $kT/\Phi=1$ (circle), 10 (square), 50 (triangle), 100 (diamond) and for $k=0$ (full line) and $k=\pi/2$ (dotted line). The bath dimension is equal to $\alpha=3$. (b) Behavior of the ratio $W^{(2)}_{k,B}(B\rightarrow F)/W^{(2)}_{k,B}(B\rightarrow B)$ with respect to the anharmonicity $A$ for the same set of parameters as in Fig 8a.  


\newpage

\begin{thebibliography}{99}

\bibitem{kn:davydov}  A. S. Davydov and N. I. Kisluka, Phys. Status Solidi \textbf{59}, 465 (1973); Zh. Eksp. Teor. Fiz \textbf{71}, 1090 (1976) [Sov. Phys. JETP \textbf{44}, 571 (1976)].
\bibitem{kn:kuznetsov} E. A. Kuznetsov and S. A. Turitsyn, Sov. Phys. JETP  \textbf{67}, 1583 (1988).
\bibitem{kn:brizhik} L. Brizhik, A. Eremko, L. Cruzeiro-Hansson and Y. Olkhovska, Phys. Rev. \textbf{B61}, 1129 (2000).
\bibitem{kn:aubry} S. Aubry, Physica \textbf{D103}, 201 (1997).
\bibitem{kn:takeno} A. J. Sievers and S. Takeno, Phys. Rev. Lett. \textbf{61}, 970 (1988).
\bibitem{kn:takeno1} S. Takeno, M. Kubota, and K. Kawasaki, Physica \textbf{D113}, 366 (1998).
\bibitem{kn:kimball} J. C. Kimball, C. Y. Fong, and Y. R. Shen, Phys. Rev. \textbf{B23}, 4946 (1981).
\bibitem{kn:scott} A. C. Scott, J. C. Eilbeck, and H. Gilhoj, Physica \textbf{D78}, 194 (1994).
\bibitem{kn:guyot1} P. Guyot-Sionnest, Phys. Rev. Lett. \textbf{67}, 2323 (1991).
\bibitem{kn:shen} R. P. Chin, X. Blase, Y. R. Shen, and S. GT. Louie, Europhys. Lett. \textbf{30}, 399 (1995).
\bibitem{kn:dai} D. J. Dai and G. E. Ewing, Surf. Sci. \textbf{312}, 239 (1994).
\bibitem{kn:jakob1} P. Jakob, Phys. Rev. Lett. \textbf{77}, 4229 (1996).
\bibitem{kn:okuyama} H. Okuyama, T. Ueda, T. Aruga, and M. Nishijima, Phys. Rev. \textbf{B63}, 233404 (2001).
\bibitem{kn:jakob2} P. Jakob, J. Chem. Phys. \textbf{114}, 3692 (2001).
\bibitem{kn:pouthier1} V. Pouthier, and C. Girardet, Phys. Rev. \textbf{B65}, 035414 (2002).
\bibitem{kn:cruzeiro1} L. Cruzeiro-Hansson, Phys. Lett. A 249 465-473 (1998); 
Phys. Lett. A 223 383-388 (1996). 
\bibitem{kn:cruzeiro2} L. Cruzeiro-Hansson and S. Takeno, Phys. Rev. \textbf{E56}, 894 (1997). 
\bibitem{kn:tekic} J. Tekic, Z. Ivic, S. Zekovic, and Z. Przulj, Phys. Rev. \textbf{E60}, 821 (1999).
\bibitem{kn:pang} Pang Xiao-feng, Eur. Phys. J. \textbf{B19}, 297 (2001).
\bibitem{kn:christiansen} P. L. Christiansen, Y. B. Gaididei, M. Johansson, and K. O. Rasmussen, Phys. Rev. \textbf{B55}, 5759 (1997).  
\bibitem{kn:jakob3} P. Jakob and B.N.J. Persson,  J. Chem. Phys. \textbf{109}, 8641 (1998).
\bibitem{kn:bogani} F. Bogani, G. Cardini, V. Schettino, and P. L. Tasselli ,  Phys. Rev. \textbf{B42}, 2307 (1990).
\bibitem{kn:pouthier2} V. Pouthier, J.C. Light, and C. Girardet, J. Chem. Phys. \textbf{114}, 4955 (2001).
\bibitem{kn:pouthier3} V. Pouthier, P. Hoang, and C. Girardet, J. Chem. Phys. \textbf{110}, 6963 (1999).
\bibitem{kn:pouthier4} V. Pouthier and C. Girardet, Phys. Rev. \textbf{B60}, 13800 (1999).
\bibitem{kn:dibartolo} B. Dibartolo, Optical interactions in Solids  (Wiley, NewYork, 1968).
\bibitem{kn:skinner}D. Hsu and J. L. Skinner, J. Chem. Phys. \textbf{81}, 1604 (1984).
\bibitem{kn:berne} S. A. Egorov, E. Rabani, and, B. J. Berne, J. Chem. Phys. \textbf{108}, 1407 (1998).
\bibitem{kn:golosov} A. A. Golosov, S. I. Tsonchev, P. Pechukas, and, R. A. Friesner, J. Chem. Phys. \textbf{111}, 9918 (1999).
\bibitem{kn:sih} R. Honke, P. Jakob, Y. J. Chabal, A. Dvorak, S.Tausendpfund, W. Stigler, P. Pavone, A. P. Mayer, and U. Schr$\ddot{o}$der, Phys. Rev. \textbf{B59}, 10996 (1999). 
\bibitem{kn:dob} L. Dobrzynski, Surf. Sci. \textbf{299/300}, 1008 (1994).

\end{thebibliography}
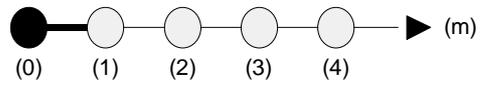
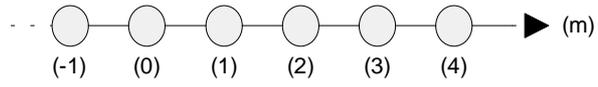
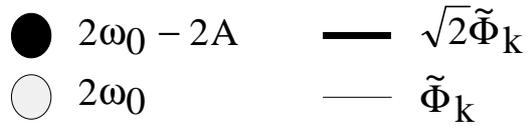
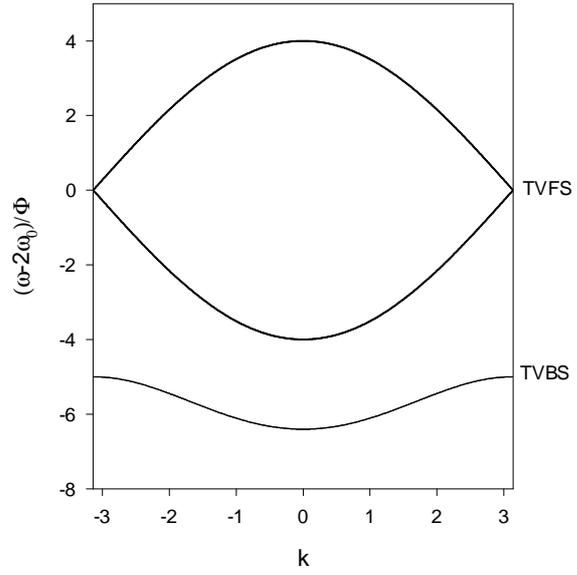
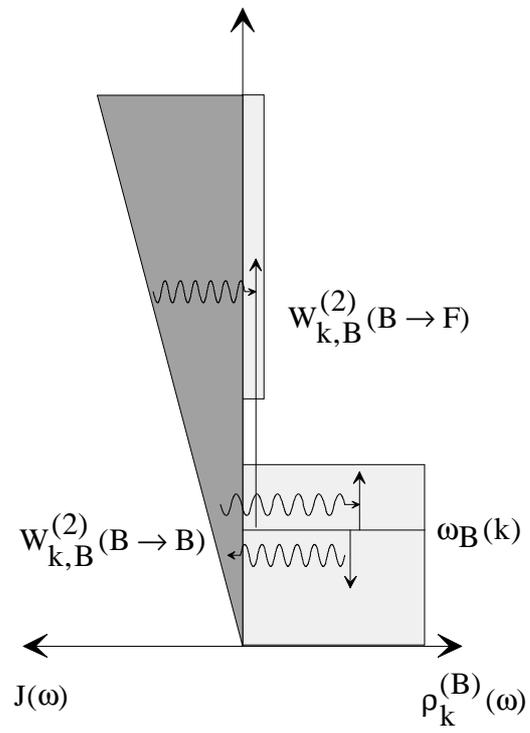
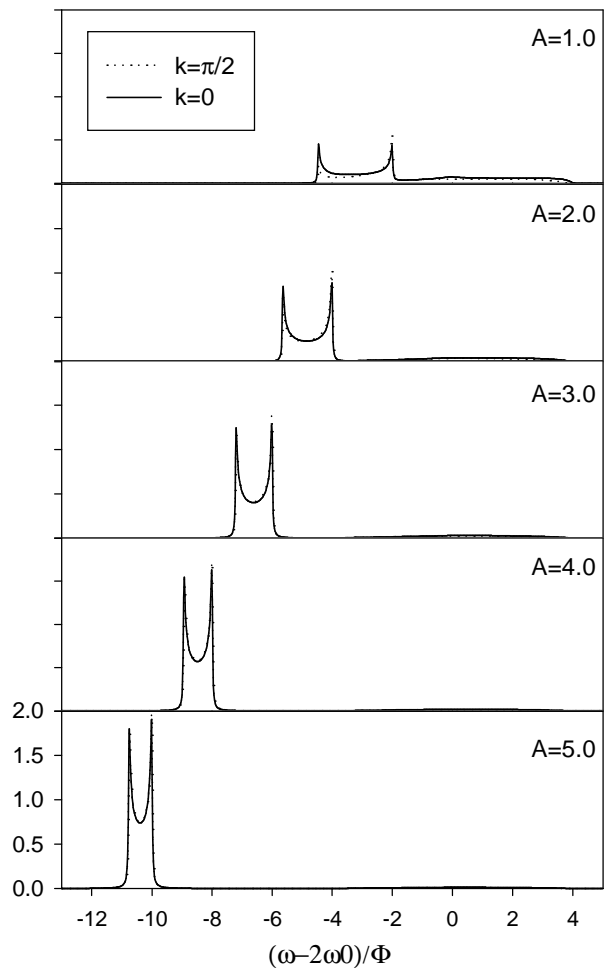
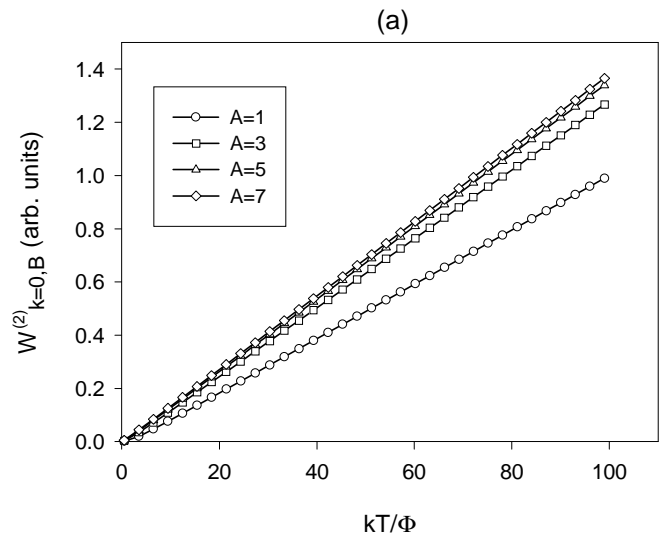
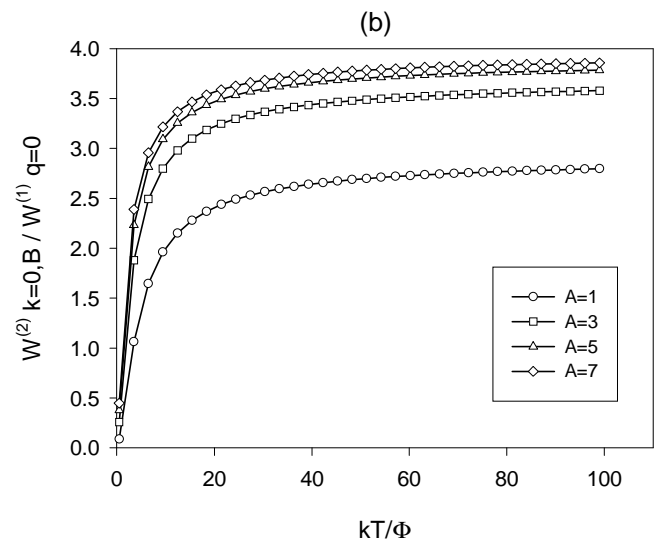
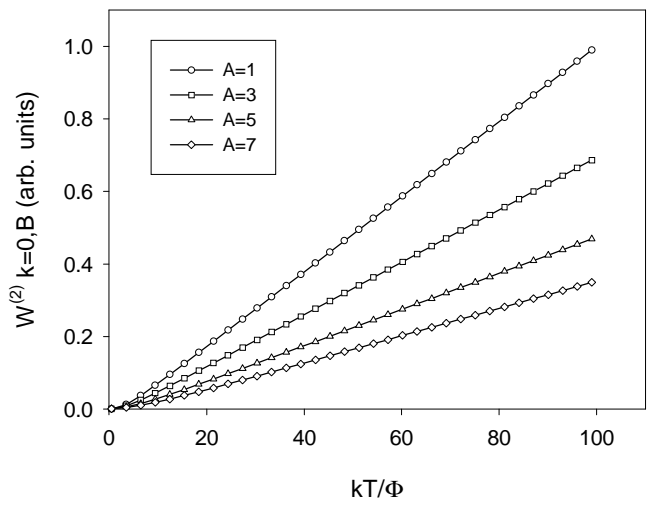
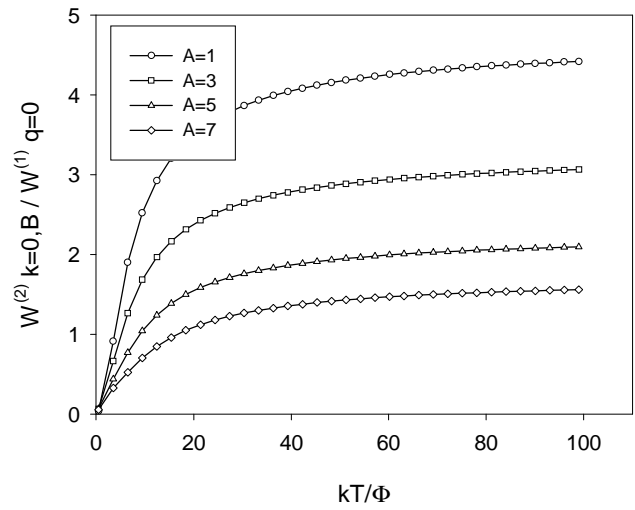
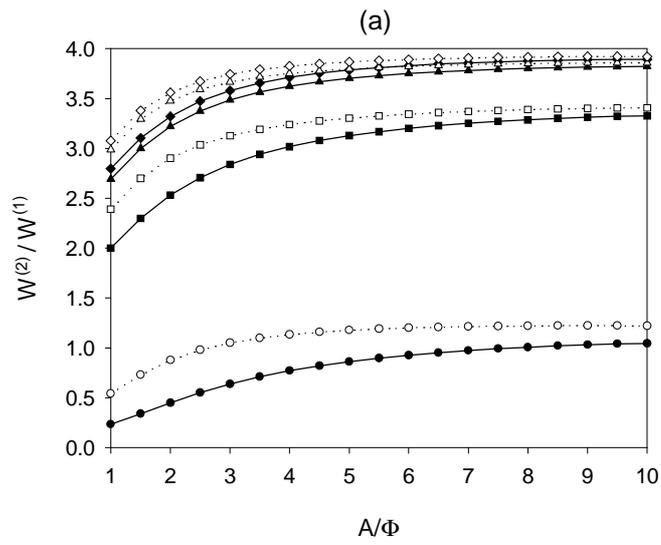
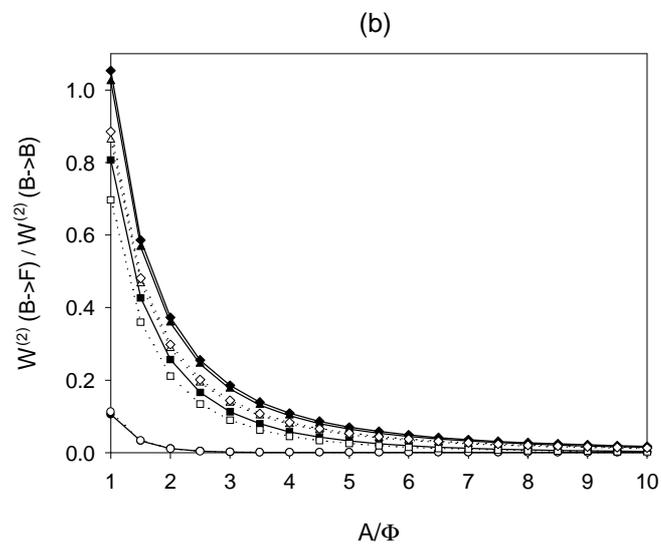
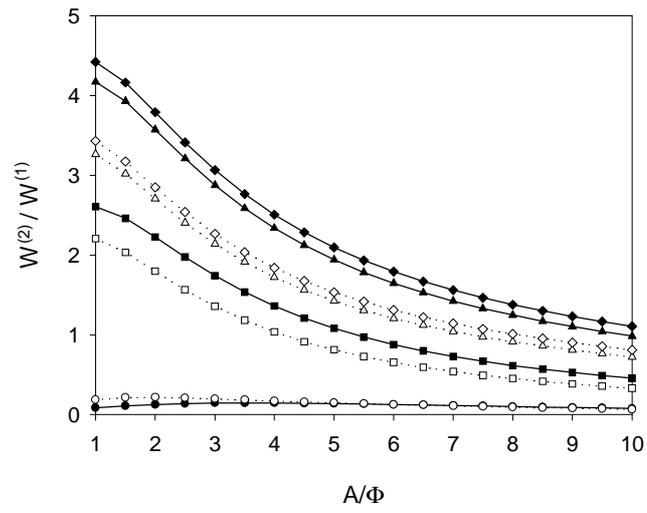
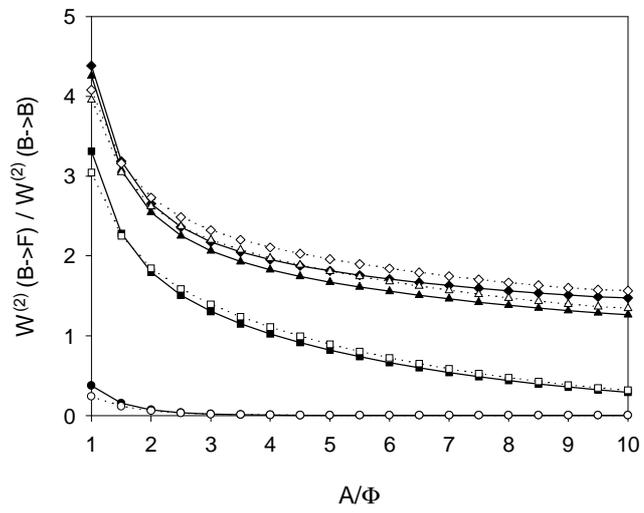
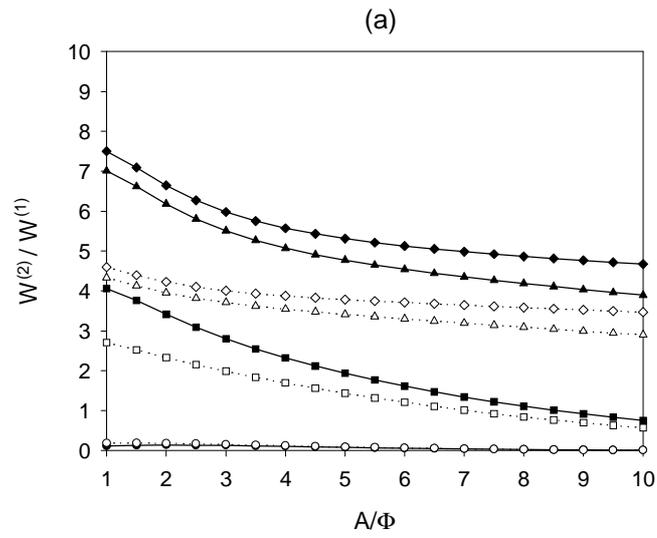
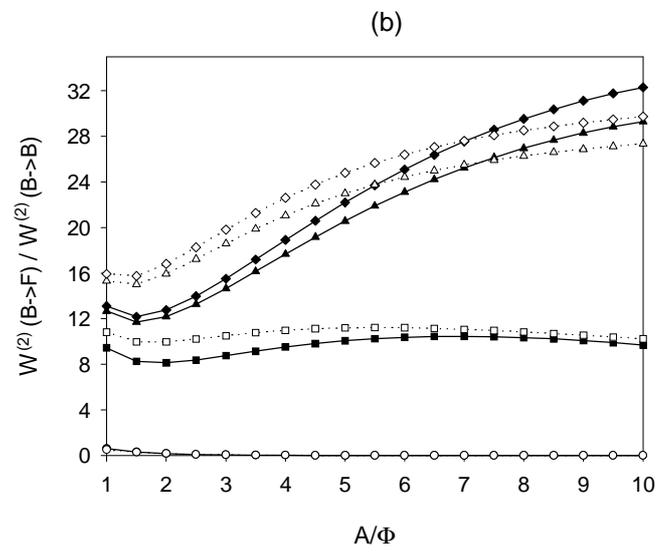
\end{document}